\documentclass[pre,aps,twocolumn,nofootinbib,floatfix]{revtex4}
\usepackage{epsf,amssymb,amsmath,multirow}
\begin{document}
\newcommand{\knn}{$\langle k_{\rm nn}\rangle (k)$}
\thispagestyle{empty}
\title{Graph theoretic analysis of protein interaction networks of eukaryotes}
\author{K.-I. Goh$^*$, B. Kahng$^{*,\dagger}$ and D. Kim$^*$}
\affiliation{$^*$School of Physics and $^{\dagger}$Program in Bioinformatics, Seoul National University, Seoul 151-747, Korea}
\date{June 15, 2004}
\begin{abstract}
Thanks to recent progress in high-throughput experimental techniques, the datasets of large-scale protein interactions of prototypical multicellular species, the nematode worm {\em Caenorhabditis elegans} and the fruit fly {\em Drosophila melanogaster}, have been assayed. The datasets are obtained mainly by using the yeast hybrid method, which contains false-positive and false-negative simultaneously. Accordingly, while it is desirable to test such datasets through further wet experiments, here we invoke recent developed network theory to test such high throughput datasets in a simple way. 
Based on the fact that the key biological processes indispensable to maintaining life are universal across eukaryotic species, and the comparison of structural properties of the protein interaction networks (PINs) of the two species with those of the yeast PIN, we find that while the worm and the yeast PIN datasets exhibit similar structural properties, the current fly dataset, though most comprehensively screened ever, does not reflect generic structural properties correctly as it is. The modularity is suppressed and the connectivity correlation is lacking. Addition of interlogs to the current fly dataset increases the modularity and enhances the occurrence of triangular motifs as well. The connectivity correlation function of the fly, however, remains distinct under such interlogs addition, for which we present a possible scenario through an {\em in silico} modeling.
\end{abstract}
\maketitle
\renewcommand{\thetable}{{\bf \arabic{table}}}
\renewcommand{\thefigure}{{\bf \arabic{figure}}}
\renewcommand{\figurename}{{\bf Fig.}} 
\renewcommand{\tablename}{{\bf Table}} 

\noindent
{\bf Introduction}\\
In the last few years graph theoretic methods to understand complex 
biomolecular systems have been developed very rapidly \cite{nrg}.
Such a development has made advances toward uncovering the organizing 
principles of cellular networks in post-genomic biology.
The cellular components such as genes, proteins, and other biological 
molecules, connected by all physiologically relevant interactions, 
form a full weblike molecular architecture in a cell. In such an 
architecture, genes play a central role, which are expressed through 
proteins. Proteins rarely act alone, rather they cooperate with others 
to act physiologically. Thus protein interactions play pivotal roles 
in various aspects of the structural and functional organizations 
and their complete description would be the first step toward a thorough 
understanding of the web of life. Proteins are viewed as 
nodes of a complex protein interaction network (PIN) in which 
two proteins are linked if they physically contact with each other.
The graph theoretic approach has been useful to understand intricate 
interwoven structures of the PIN \cite{jeong,wagner,maslov}. 
The key biological processes indispensable to maintaining life
are universal across eukaryotic species since many involved genes 
are evolutionarily conserved \cite{alberts}. 
Using this property, one can test a newly discovered dataset 
if it really contains more or less complete information of protein 
interactions. Moreover, this {\it in silico} approach offers 
one the candidates of protein interaction pairs, of which the number 
is considerably reduced compared with the total combinatorial 
pairs. Thus, the graphic theoretic analysis would provide a useful 
guide for further wet studies of protein interactions.

Species with sequenced genome such as the yeast {\em Saccharomyces 
cerevisiae} provide important test beds for the study of the PIN. 
Thanks to recent progress in the high-throughput experimental 
techniques such as the yeast two-hybrid assay \cite{uetz,ito} and 
the mass spectroscopy \cite{gavin,ho},  
the dataset of the yeast PIN has been firmly established~\cite{mips,dip}.
Very recently, large-scale protein interactions of multicellular species, 
the nematode worm {\em Caenorhabditis elegans} \cite{vidal} 
and the fruit fly {\em Drosophila melanogaster} \cite{giot}, 
have been assayed. While those datasets, mainly based on the yeast 
two-hybrid assay, need physiological proof, they contain large-scale 
proteins and protein interactions, making graph theoretic study possible. 
In this paper, we analyze those datasets and compare them with the 
more-established set of interactions in the budding yeast~\cite{dip}. 
Our graph theoretic analysis suggests that the present interaction dataset 
of the fruit fly, based on the yeast two-hybrid (Y2H) assay, may have left out 
a significant part of protein interactions, though most comprehensively 
screened ever. Such conclusion has been reached by the comparison of 
the generic features of the PIN, the modularity and the 
connectivity correlations, across the three species. For the fly, those 
quantities behave distinctively: The modularity is suppressed 
and the connectivity correlation is lacking. Such distinct behavior 
can be overcome partially by the addition of yeast interlogs into the 
fly dataset.\\

\begin{figure*}[t]
\centerline{\epsfxsize=15cm \epsfbox{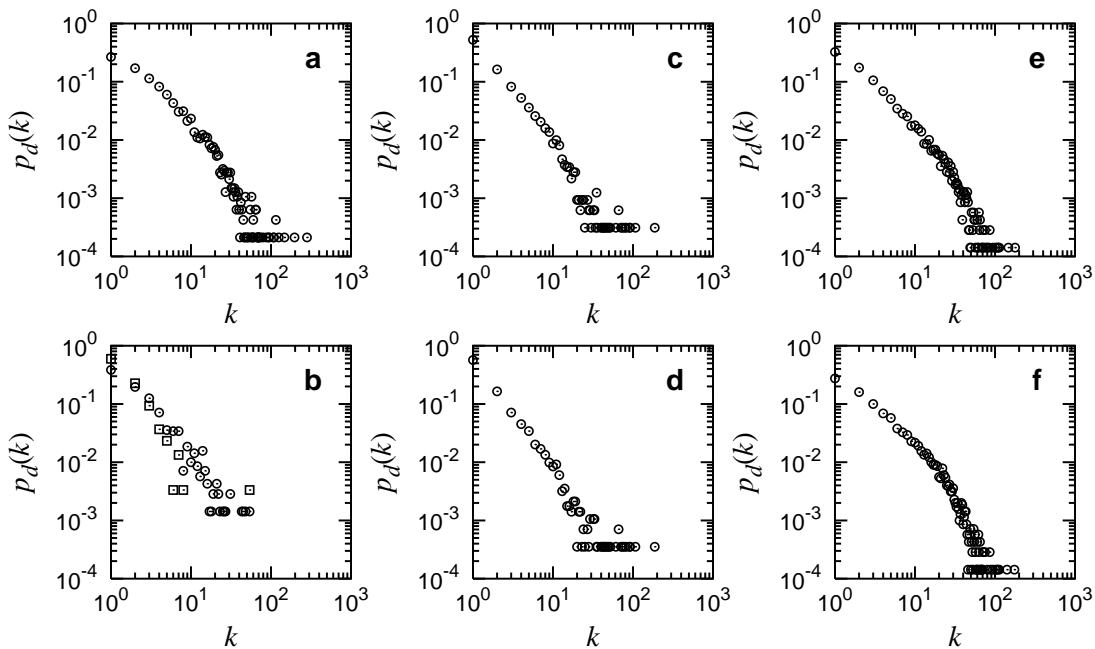}}
\caption{The degree distributions $p_d(k)$ for
{\bf (a)} the yeast, 
{\bf (b)} the prokaryotes {\em Helicobacter pylori} ($\circ$)
and {\em Escherichia coli} ($\Box$),
{\bf (c)} the worm (Worm-All),
{\bf (d)} the Y2H subset of the worm dataset (Worm-Y2H),
{\bf (e)} the fly,
and {\bf (f)} the Fly+Interlog dataset.
}
\label{fig:pk}
\end{figure*}

\noindent{\bf Materials and Methods}\\
{\bf Graph theory terminology.}
(i) Network is composed of vertices and edges. In the 
protein interaction network, vertices represent proteins and 
edges protein interactions. 
(ii) Degree is the number of edges connected to a given 
vertex. The degree distribution $p_d(k)$ is the fraction of vertices 
having $k$ degrees.
(iii) Clustering coefficient of a node is 
defined as $C_i=2e_i/k_i(k_i-1)$, where $e_i$ is the number 
of connections among the $k_i$ neighbors of a vertex $i$. 
Clustering function $C(k)$ is the mean value of $C_i$ over 
the vertices with degree $k$, while the clustering coefficient
$C$ is the mean of $C_i$ over all vertices.  
When the network contains 
hierarchical and modular structures within it, it is known 
that the clustering function $C(k)$ behaves as 
$C(k)\sim k^{-\beta}$ for large $k$~\cite{ravasz}.  
(iv) $\langle k_{\rm nn} \rangle(k)$ is the mean degree 
of the neighbors of a vertex with degree $k$. 
It is known that $\langle k_{\rm nn} 
\rangle (k)\sim k^{-\nu}$ with $\nu > 0$ for the Internet and 
the protein interaction network~\cite{knn,maslov}, implying that
vertices with large degree tend to connect to the ones with small 
degree. Such a network is called dissortative network. 
Besides this quantity, the ep x $r$ has been 
introduced \cite{assort} to characterize the degree-degree 
correlation between the two vertices located at the ends of 
an edge, which is defined as     
\begin{equation*}
r = \frac{\langle k_1k_2\rangle - \langle (k_1+k_2)/2\rangle^2}{
\langle (k_1^2+k_2^2)/2\rangle - \langle (k_1+k_2)/2\rangle^2} \quad,
\end{equation*}
where $k_1$ and $k_2$ are the degrees of two vertices at the ends of
an edge, and $\langle \cdots\rangle$ denotes the average over all
edges.\\

\noindent
{\bf The protein interaction network datasets.}
We used the yeast subset of the interaction data compiled in the 
Database of Interacting Proteins (DIP) as of January 2004 
(http://dip.doe-mbi.ucla.edu) \cite{dip}. The datasets for the 
worm and the fly are obtained from the works of Li {\em et al.}~\cite{vidal}
and Giot {\em et al.}~\cite{giot}, 
respectively. 
For the worm, we consider two different versions, the one consisting
of only the interactions from the Y2H screens (referred to as Worm-Y2H
network in this paper) 
and the other the full network supplied by Li {\em et al.}~\cite{vidal}
(referred to as Worm-All network).
The characteristics of each dataset and the values 
of the graphic theoretic quantities are tabulated in Table~\ref{table1}.\\

\begin{table}[b]
\caption{
{\bf Protein interaction network datasets.}
Tabulated are for each dataset the size of proteome $N_{\rm proteome}$, 
the number of proteins $N$ and the number of protein--protein
interactions $L$ in the dataset, the mean degree $\langle k\rangle$,
the clustering coefficient $C$, the assortativity $r$,
and the number of proteins forming the largest cluster $N_1$.
The self-interactions are excluded throughout.
}
\begin{ruledtabular}
\begin{tabular}{cccccc}
 & Yeast & Worm-Y2H & Worm-All & Fly \\
\hline
$N_{\rm proteome}$ & 6195 & 22246 & 22246 & 16206\\
$N$ & 4714 & 2835 & 3216 & 7055 \\
$L$ & 14857 & 4438 & 50444 & 20947\\
$\langle k\rangle$ & 6.3 & 3.1 & 3.4 & 5.9\\
$C$ & 0.12 & 0.047 & 0.15 & 0.014\\
$r$ & -0.14 & -0.16 & -0.13 & -0.036\\
$N_1$ & 4627 & 2601 & 2898 & 6929
\label{table1}
\end{tabular}
\end{ruledtabular}
\end{table}

\noindent
{\bf Orthologous gene assignment.}
For cross-species ortholog information, we used the information from the 
KOG database \cite{kog}, a eukaryotic extension of the Clusters of 
Orthologous Genes (COG) database (http://www.ncbi.nlm.nih.gov/COG/new/).

\begin{figure*}[t]
\centerline{\epsfxsize=15cm \epsfbox{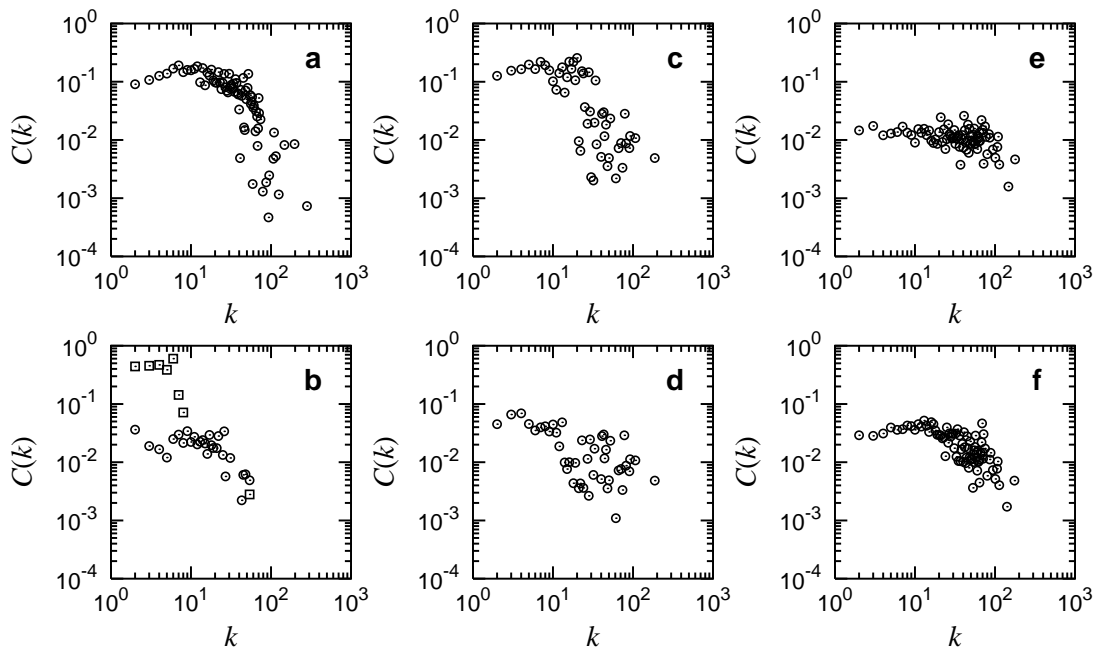}}
\caption{The local clustering function $C(k)$ for
{\bf (a)} the yeast, 
{\bf (b)} the bacteria {\it H. pylori} ($\circ$) and {\em E. coli} ($\Box$), 
{\bf (c)} the worm (Worm-All),
{\bf (d)} the Worm-Y2H dataset, 
{\bf (e)} the fly, and 
{\bf (f)} the Fly+Interlog dataset.
The abscissae and ordinates are fixed for clear comparison.
}
\label{fig2-ck}
\end{figure*}

\noindent
{\bf Yeast interlogs in fly.}
Having identified the yeast-fly orthologs, we look for the 
interactions in the yeast network between
those yeast proteins both having orthologs in the fly network.
Such orthologous interactions are called the interlogs.
If the corresponding fly interaction is present, we call
it an {\em overlap interlog}. If not, we call it a {\em potential
interlog}. Note that the ortholog relationship is not always one-to-one,
resulting in multiple interlogs for a given yeast interaction.
For {\em in silico} analysis on the effect of the addition of
potential interlogs in the fly network, we include on average one potential
interlog per yeast interaction. Specifically, for each
yeast interaction A-B having no overlap interlog, each potential interlog
is added in the fly network with probability $1/(o_Ao_B)$,
where $o_X$ is the number of fly ortholog(s) of the yeast gene X. 
The network obtained in this way
is referred to as Fly+Interlog network hereafter.
The full list of the 408 overlap and the 55176 potential interlogs 
are available on the web 
(http://komplex0.snu.ac.kr/pin/yeast-fly-interlog.xls).
\\

\noindent{\bf Results}\\
\noindent{\bf Degree distributions.} 
In Fig.~\ref{fig:pk}, we plot the degree distributions of diverse 
protein interaction networks, all of which display the scale-free
behavior, fitting well to the generalized Pareto formula, 
$p_d(k)\sim (k+k_0)^{-\gamma}$,
almost indistinguishable with each other. 
While the degree distribution is a fundamental quantity in graph theory,
it deals with global network structure, so it does not give 
detailed information on structural property.
\\

\noindent{\bf Modularity}\\ 
A cellular function is achieved by a set of related
proteins, usually forming a pathway or a complex.
Such functional module manifests itself as a localized dense
subgraph within the whole cellular network.
The presence of modules and their hierarchical organization
can be visualized by the local clustering function $C(k)$ \cite{ravasz}. 
For the yeast PIN, $C(k)$ exhibits a plateau
for small $k$ and falls off rapidly for large $k$, 
reflecting the modular structure bridged by the hubs (Fig.~2{\bf a}).
The similar pattern is observed in the worm (Fig.~2{\bf c}) 
and the two prokaryotic species, {\em H. pylori} 
and {\em E. coli} (Fig.~2{\bf b}). 
Note that the worm dataset contains the yeast interlogs. 
For the fly Y2H data, however, $C(k)$ behaves distinctively,
almost constant for all $k$ (Fig.~2{\bf e}).  
To understand this discrepancy, we add
the potential yeast interlogs into the current fly Y2H dataset. 
Then $C(k)$ behaves in a similar fashion to other dataset, 
showing a moderate plateau for small $k$ and rapid decrease for 
large $k$, albeit the altitude of the plateau, which is roughly
the clustering coefficient $C$, is not as high as in the yeast and
the worm (Fig.~2{\bf f}).
To find the role of the interlogs in the worm, we consider
the Worm-Y2H dataset, and plot its $C(k)$ in Fig.~2{\bf d}. 
Indeed, the signature character of $C(k)$ 
is lost, in particular, the plateau for small $k$ almost disappears, 
implying  the yeast-interlogs play a role of forming modules, where 
proteins are closely linked each other.\\       

\noindent{\bf Conservation rate of interactions.} 
We count how many yeast interactions are actually conserved 
in orthologous form in both the worm and the fly. 
The conservation rate found in this way for the Y2H screen dataset
is surprisingly low; 2.7\% for the worm (Worm-Y2H)
and 3.8\% for the fly. 
For the worm, we note that such low coverage is in part 
due to the insufficient number of baits used in the experiment 
(3,024 baits, 833 out of which are present in the network).
When we consider the conservation of triangular interaction patterns,
a basic unit of cooperative functional module \cite{milo},
only 3 out of 1731 are conserved in the worm, while none in the fly 
(Fig.~3). 
The lack of conserved interaction motifs in the fly data
suggests that the current fly network misses some of important 
cooperative aspects of the cellular network in the fly.
The effort to fill this gap is timely.\\

\begin{figure}
\centerline{\epsfxsize=8.5cm \epsfbox{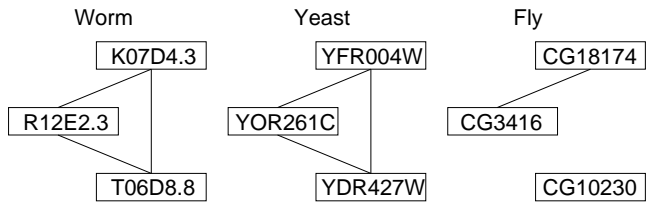}}
\caption{
Conservation of interaction motif.
Shown in the middle is a triangular interaction subgraph within the yeast involving in 
ubiquitin-dependent protein catabolism. Corresponding
orthologous counterpart in the worm and the fly are also shown.
This motif is conserved in the worm Y2H data, while only a single interaction
is detected in the fly data.
}
\end{figure}

\begin{table*}
\caption{Network motif structure of the three species.
Tabulated is the number of each subgraph present in the network. 
According to its $Z$- and $E$-score, the significant motifs (M) and 
anti-motifs (AM) are indicated.}
\label{motif-table}
\begin{tabular}{cccccc}
 & Yeast & Worm-Y2H & Worm-All & Fly & Fly+Interlog \\
\hline
\multirow{2}{5mm}{
\begin{picture}(0,0)(0,0)
 \qbezier(0,0)(4,6.8)(4,6.8)
 \qbezier(4,6.8)(4,6.8)(8,0)
 \put(0.2,0){\circle*{3}}
 \put(8.2,0){\circle*{3}}
 \put(4.2,6.8){\circle*{3}}
\end{picture}
}
& 329961 & 81205 & 87294 & 413926 & 520704$\pm$1358 \\
&  &  &  &  & \\
\hline
\multirow{2}{5mm}{
\begin{picture}(0,0)(0,0)
 \qbezier(0,0)(4,6.8)(4,6.8)
 \qbezier(4,6.8)(4,6.8)(8,0)
 \qbezier(0,0)(0,0)(8,0)
 \put(0.2,0){\circle*{3}}
 \put(8.2,0){\circle*{3}}
 \put(4.2,6.8){\circle*{3}}
\end{picture}
}
& 7136 & 366 & 1512 & 1549 & 3504$\pm$40 \\
& M (Z=80, E=3.3) & & M (Z=29, E=2.5) & & M (Z=45, E=1.4)\\
\hline
\multirow{2}{5mm}{
\begin{picture}(0,0)(0,0)
 \qbezier(0,0)(0,8)(0,8)
 \qbezier(0,8)(8,8)(8,8)
 \qbezier(8,8)(8,0)(8,0)
 \put(0.2,0){\circle*{3}}
 \put(8.2,0){\circle*{3}}
 \put(0.2,8){\circle*{3}}
 \put(8.2,8){\circle*{3}}
\end{picture}
}
& 4081023 & 604723 & 680485 & 7378808 & 971960$\pm$37157 \\
&  &  &  &  &  \\
\hline
\multirow{2}{5mm}{
\begin{picture}(0,0)(0,0)
 \qbezier(0,0)(0,8)(0,8)
 \qbezier(0,8)(8,8)(8,8)
 \qbezier(0,8)(8,0)(8,0)
 \put(0.2,0){\circle*{3}}
 \put(8.2,0){\circle*{3}}
 \put(0.2,8){\circle*{3}}
 \put(8.2,8){\circle*{3}}
\end{picture}
}
& 9024723 & 2129609 & 2157048 & 6315922 & 7409320$\pm$24476 \\
&  &  &  &  &  \\
\hline
\multirow{2}{5mm}{
\begin{picture}(0,0)(0,0)
 \qbezier(0,0)(0,8)(0,8)
 \qbezier(0,8)(8,8)(8,8)
 \qbezier(0,0)(0,0)(8,0)
 \qbezier(0,8)(8,0)(8,0)
 \put(0.2,0){\circle*{3}}
 \put(8.2,0){\circle*{3}}
 \put(0.2,8){\circle*{3}}
 \put(8.2,8){\circle*{3}}
\end{picture}
}
& 368730 & 46050 & 58520 & 160846 & 263324$\pm$2617 \\
& AM (Z=-122, E=-0.7) &  & AM (Z=-59, E=-0.7) &  &  \\
\hline
\multirow{2}{5mm}{
\begin{picture}(0,0)(0,0)
 \qbezier(0,0)(0,8)(0,8)
 \qbezier(0,8)(8,8)(8,8)
 \qbezier(8,8)(8,0)(8,0)
 \qbezier(8,0)(8,0)(0,0)
 \put(0.2,0){\circle*{3}}
 \put(8.2,0){\circle*{3}}
 \put(0.2,8){\circle*{3}}
 \put(8.2,8){\circle*{3}}
\end{picture}
}
& 21806 & 4350 & 4686 & 54100 & 60648$\pm$206 \\
&  & M (Z=9.5, E=0.6) &  & M (Z=81, E=2.1)  & M (Z=60, E=1.3) \\
\hline
\multirow{2}{5mm}{
\begin{picture}(0,0)(0,0)
 \qbezier(0,0)(0,8)(0,8)
 \qbezier(0,8)(8,8)(8,8)
 \qbezier(8,8)(8,0)(8,0)
 \qbezier(8,0)(8,0)(0,0)
 \qbezier(0,8)(8,0)(8,0)
 \put(0.2,0){\circle*{3}}
 \put(8.2,0){\circle*{3}}
 \put(0.2,8){\circle*{3}}
 \put(8.2,8){\circle*{3}}
\end{picture}
}
& 27455 & 1505 & 4120 & 4029 & 9313$\pm$228 \\
& AM (Z=-49, E=-0.7) &  & AM (Z=-25, E=-0.6) & M (Z=12, N=0.8) &  \\
\hline
\multirow{2}{5mm}{
\begin{picture}(0,0)(0,0)
 \qbezier(0,0)(0,8)(0,8)
 \qbezier(0,8)(8,8)(8,8)
 \qbezier(8,8)(8,0)(8,0)
 \qbezier(8,0)(8,0)(0,0)
 \qbezier(0,8)(8,0)(8,0)
 \qbezier(8,8)(0,0)(0,0)
 \put(0.2,0){\circle*{3}}
 \put(8.2,0){\circle*{3}}
 \put(0.2,8){\circle*{3}}
 \put(8.2,8){\circle*{3}}
\end{picture}
}
& 5259 & 30 & 1563 & 82 & 914$\pm$35 \\
&  &  & M (Z=10, E=0.8) & M (Z=11, E=3.5) &  M (Z=40, E=6.0)\\
\hline
\end{tabular}
\end{table*}

\noindent{\bf Motif structure.}
Since the modularity manifested by $C(k)$ 
is closely related to the formation of triangles in the network,
here we further perform network motif analysis
for the three species datasets. 
The network motifs are small recurring subgraphs which are overrepresented
in a given network and are believed to provide the basic evolutionary
and functional signatures of the network \cite{milo}.
Since it was recently discovered that the motif constituents 
are more conserved during evolution than the rest \cite{wuchty}, 
one would expect the density of each motif to be 
close to each other across the three species. 
From the comparison of the columns for Yeast, Worm-All, and Fly 
in Table \ref{motif-table}, we can see that 
the triangle motif is relatively not abundant in Fly, 
while the square motif is. 
Thus, the absolute magnitude of the clustering function 
is smaller for the fly than for the yeast or the worm. 
The density of the triangle motif is higher in the Fly+Interlog dataset, 
indicating that the clustering coefficient is enhanced overall by the 
addition of the interlogs of the fly.

In Table \ref{motif-table} we have summarized the motif structure
for each network.
We follow Milo {\em et al.}~\cite{milo} to calculate the two scores, 
$Z$- and $E$-score, defined as
$Z=(N-N_{random})/\sigma_{random}$ and $E=(N-N_{random})/N_{random}$, 
respectively, and use the following two criteria to specify whether a
subgraph is a motif or an anti-motif (an anti-motif is a subgraph
significantly underrepresented in the network):
\begin{itemize}
\item[(i)] The probability that $N$ is observed in randomized network 
is smaller than 0.01. 
\item[(ii)] $|E|>E_0$, where we set the threshold $E_0=0.5$, 
rather than $E_0=0.1$ in Milo {\em et al.}~\cite{milo}.
\end{itemize}
Here, $N_{random}$ and $\sigma_{random}$ are the expected number of occurrence
in the randomized version of the network and their standard deviation 
obtained from 1000 samples respectively, 
where the randomization is performed by the 
switching method \cite{milo}. In calculating them for 
the 4-node subgraphs, the numbers of 3-node subgraphs are fixed to be 
those of the original networks. For the Fly+Interlog network,
10 realizations of interlog addition (see Method) are averaged. 
\\

\begin{figure*}
\centerline{\epsfxsize=15cm \epsfbox{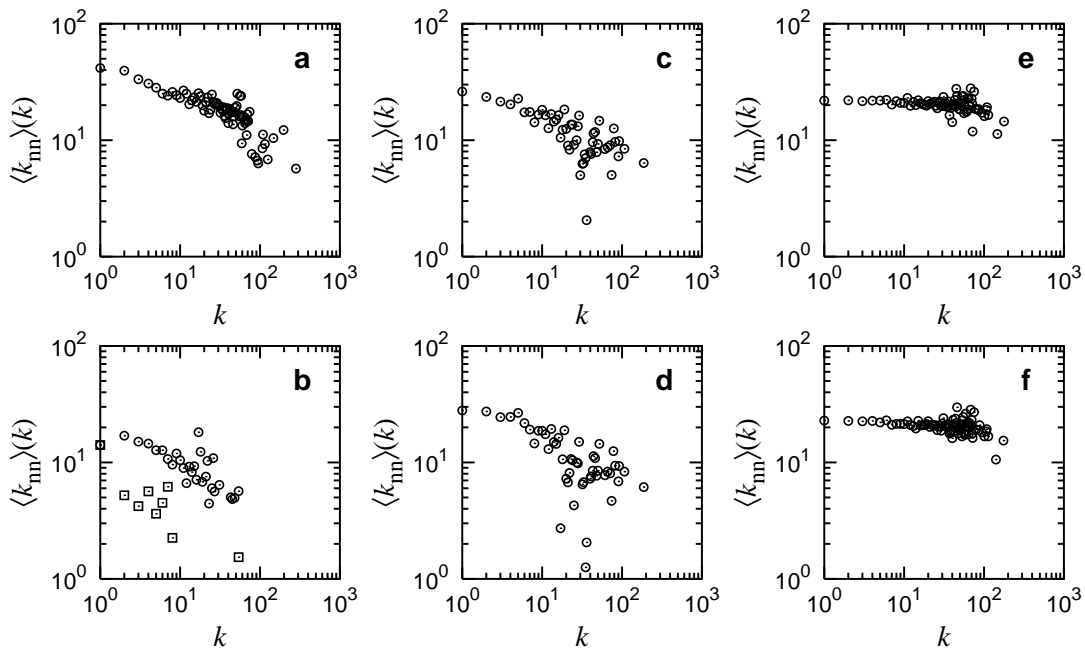}}
\caption{
The average neighbor degree function $\langle k_{\rm nn}\rangle(k)$ for
{\bf (a)} the yeast,
{\bf (b)} the prokaryotes {\em H. pylori} ($\circ$) and {\em E. coli} ($\Box$),
{\bf (c)} the worm (Worm-All),
{\bf (d)} the Y2H subset of the worm (Worm-Y2H),
{\bf (e)} the fly,
and {\bf (f)} the Fly+Interlog dataset.
The abscissae and ordinates are fixed for clear comparison.
}
\label{fig-knn}
\end{figure*}

\noindent{\bf Degree-degree correlation}\\
The mean neighbor degree function \knn~is useful 
in understanding the degree-degree correlation in a network. 
In Fig.~\ref{fig-knn}, we plot \knn~for each dataset.
For the yeast, it is known that \knn~decreases 
with increasing $k$ \cite{maslov}, which turns out to be also true
for some prokaryotic species, too (Figs.~4{\bf a}-{\bf b}). 
Such a behavior in \knn~is also observed for the worm 
(Figs.~4{\bf c}-{\bf d}), however, 
it is flat for the fly, implying lack of correlation (Fig.~4{\bf e}). 
Such distinct behavior for the fly is robust under the addition of
the interlogs (Fig.~4{\bf f}), which suggests the lack of 
correlation in the fly network could be intrinsic,
even though we cannot exclude the possibility 
that it is again the artifact of the incompleteness of the data.
The hypothesis that the lack of correlation could be intrinsic 
may be supported by the following observations.\\

\noindent
{\bf Effect of diversification of gene function on $\langle k_{\rm nn}\rangle(k)$.}
While the pattern of $C(k)$ of the fly becomes similar to those 
of the yeast and the worm by the addition of the interlogs, 
that of $\langle k_{\rm nn} \rangle(k)$ remains distinct.
Thus here we investigate if such a flat behavior is intrinsic 
through an {\it in silico} model, finding that indeed, 
the decreasing behavior of $\langle k_{\rm nn}\rangle(k)$ 
becomes moderated through the network evolution with the duplication 
and divergence processes. Homologs in a genome 
are thought to result from the gene duplication event, which is 
usually followed by the diversification to lower the redundancy. 
Some computer models aiming to mimic these processes
in proteome evolution exist in the literature \cite{sole,vespig}.
We investigate how the diversification process
affects the topological property of the proteome network, 
in particular, the degree-degree correlation in terms of 
$\langle k_{\rm nn} \rangle(k)$. 
To this end, we perform following procedures motivated by
V\'azquez {\em et al.}~\cite{vespig}:
\begin{enumerate}
\item Starting with the yeast protein network, at each step,
a protein A is chosen randomly and is duplicated as A$'$.
Then the protein A and A$'$ share common neighbors.
\item For each neighboring protein of A and A$'$, one of edges connected
to either A or A$'$ is removed with equal probability.
\item Repeat 1--2 until the number of proteins reaches $\sim$20,000,
the approximate sizes of the worm and the fly proteome.
\end{enumerate}
Note that in this procedure, the number of proteins increases while
the number of interactions stays still. Thus the average degree 
decreases as the size of proteome increases. Such decrease
will be compensated by, e.g., the acquisition of new interactions
between existing proteins via mutation. However, we do not take
such a process into account, to single out the effect of the
diversification only.

\begin{figure}[!t]
\centerline{\epsfxsize=9cm \epsfbox{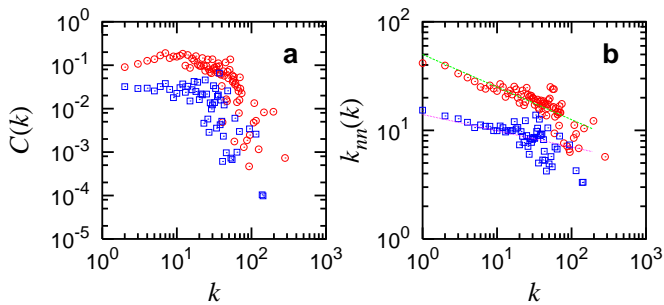}}
\caption{Effect of gene function diversification in (a) $C(k)$ and (b)
\knn. Red circles are the data
of the original yeast network and the blue squares those after
running the diversification procedures {\it in silico}. 
The slope of the straight line
(the rate of decrease) in (b) is -0.3 (top, green) 
and -0.15 (bottom, magenta), respectively.}
\label{model}
\end{figure}

The result of simulation is shown in Fig.~\ref{model}.
The local clustering function $C(k)$ is simply shifted downward, due to
the overall decrease of the edge density.
On the other hand, the average neighbor degree 
\knn decreases as $k$ but with a smaller rate, indicating that the 
diversification process can, although not perfectly, neutralize 
the connectivity correlation.
Furthermore, if we {\em assume} that the establishment of new interactions
follows the preferential attachment \cite{ba} or random attachment,
the overall correlation would diminish eventually.\\

\begin{figure*}
\centerline{\epsfxsize=15cm \epsfbox{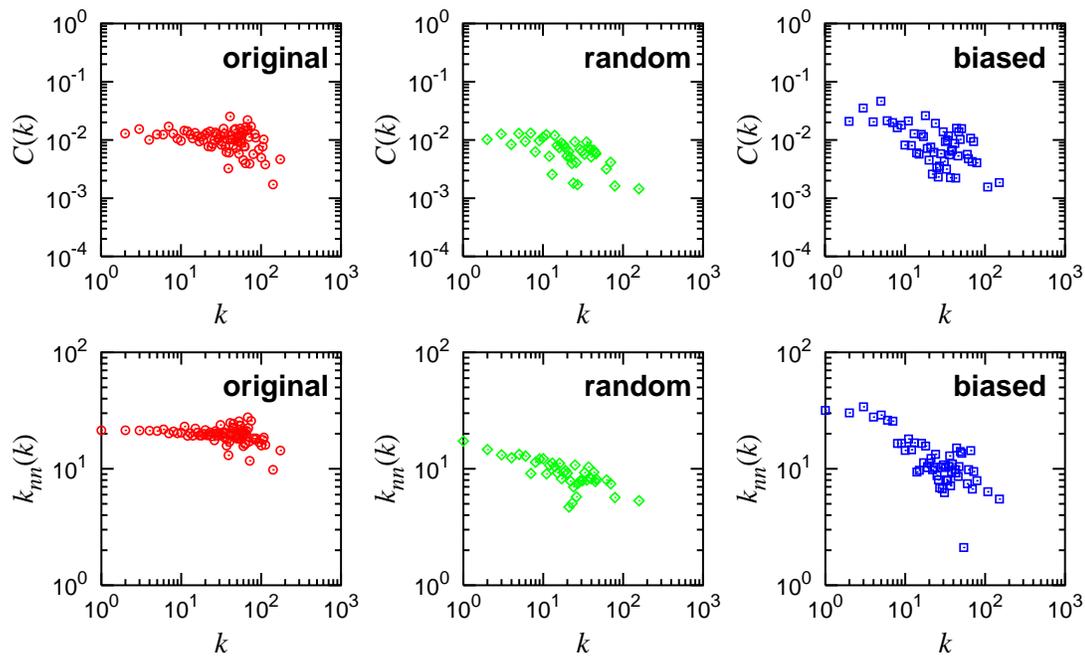}}
\caption{Effect of bait selection. Red circle is for the full data,
green diamond the randomly sampled one, blue square the biased sampled one.}
\label{bait}
\end{figure*}

\noindent
{\bf Effect of bait selection on $\langle k_{\rm nn}\rangle(k)$.}
There has been an argument that the apparent decreasing trend
in $\langle k_{\rm nn}\rangle(k)$ is an artifact from the limited 
selection of baits in the two-hybrid experiment \cite{aloy}.
Indeed, Li et al.~\cite{vidal} had selected the baits with their 
own criteria, mainly based on the biological indispensability
and the potential applicability to the human therapeutics.
To check this hypothesis {\it in silico}, we sampled the 30\% subset
of 4950 baits identified in Giot et al.'s fly network \cite{giot} and 
reconstructed the network only with the interactions associated with 
the sampled baits.
We sampled in two different ways; the random sampling
and the biased sampling toward the highly connected 
baits (the sampling probability is proportional to
the number of bait-interactions).
Both data sets generate the decreasing trend 
in $\langle k_{\rm nn}\rangle(k)$ (Fig.~\ref{bait}).
One can see that even though the original network has the null slope in 
$\langle k_{\rm nn}\rangle (k)$,
the negative slope develops in the sampled ones, demonstrating 
that the insufficient use of the bait {\em can} produce 
artifactual correlation
in the connectivity. If this scenario holds, one conjecture that
$\langle k_{\rm nn} \rangle(k)$ curve will become flatter
as the interaction data accumulates and becomes more complete.
\\

\noindent{\bf Summary and discussion}\\
We have investigated in detail the structural properties
of the protein interaction networks of three eukaryotic species,
the budding yeast, the nematode worm, and the fruit fly.
In particular, we have focused on the comparative assessment of
the modularity and the degree-degree correlation for those networks.
We found that while the worm dataset behaves similarly to the 
yeast for the two graph theoretic quantities, the fly does not. 
The difference might be attributed to the presence (absence) of 
the yeast-interlogs in the current worm (fly) dataset.  
For the fly dataset, the modularity is suppressed and 
the connectivity correlation is lacking. We found that 
the clustering function can be restored to those of the yeast 
dataset by the addition of interlogs selected randomly 
among the candidates to the current dataset. 
We also performed motif analysis for the three species, finding 
that the density of the triangle motif is increased by the 
addition of the interlogs to the current fly dataset. 
Finally, the candidates of the protein interactions of the fly 
are provided in the supplementary materials, 
which could be useful in finding protein interactions missed in 
the current fly dataset. 
\\

This work is supported by the KOSEF grant No. R14-2002-059-01000-0
in the ABRL program and the MOST grant No. M1 03B500000110.

\end{document}